\documentstyle[11pt]{article}
\begin{document}

\title{Rotating Bianchi type $V$ dust models generalizing the $k = -1$ Friedmann
model}

\author{Andrzej Krasi\'nski\thanks{This research was supported by the Polish
Research Committee grant no 2 P03B 060 17.}}

\date {}

\maketitle

\centerline{N. Copernicus Astronomical Center, Polish Academy of Sciences }

\centerline{Bartycka 18, 00 716 Warszawa, Poland}

\centerline{email: akr@camk.edu.pl}

\begin{abstract}
The Einstein equations for one of the hypersurface-homogeneous rotating dust
models are investigated. It is a Bianchi type $V$ model in which one of the
Killing fields is spanned on velocity and rotation (case 1.2.2.2 in the
classification scheme of the earlier papers). A first integral of the field
equations is found, and with a special value of this integral coordinate
transformations are used to eliminate two components of the metric. The $k =
-1$ Friedmann model is shown to be contained among the solutions in the limit
of zero rotation. The field equations for the simplified metric are reduced to
3 second-order ordinary differential equations that determine 3 metric
components plus a first integral that algebraically determines the fourth
component. First derivatives of the metric components are subject to a
constraint (a second-degree polynomial with coefficients depending on the
functions). It is shown that the set does not follow from a Lagrangian of the
Hilbert type. The group of Lie point-symmetries of the set is found, it is
two-dimensional noncommutative. Finally, a method of searching for first
integrals (for sets of differential equations) that are polynomials of degree 1
or 2 in the first derivatives is applied. No such first integrals exist. The
method is used to find a constraint (of degree 1 in first derivatives) that
could be imposed on the metric, but it leads to a vacuum solution, and so is of
no interest for cosmology.
\end{abstract}

\section{Statement of the problem and summary of the paper.}
\par
\setcounter{equation}{0} This paper is a continuation of a series of papers on
rotating dust models in relativity$^{1-3}$. The initial motivation for this
research was the desire to find a rotating generalization of the Friedmann
models. In spite of much effort spent on investigating solutions of Einstein
equations with a rotating matter source, no such generalization has been found
so far; see literature surveys in Refs. 3 and 4. Refs. 1, 2 and 3 provided a
complete classification scheme for hypersurface-homogeneous rotating perfect
fluid models with zero acceleration. Unlike in previous approaches, nothing was
assumed about the position of the symmetry orbits in spacetime; the
classification includes also timelike and null orbits, and so it is the
farthest-reaching application of the Bianchi classification to rotating and
nonaccelerating perfect fluid models in relativity. The models split into 3
general classes: I, in which two of the Killing fields are everywhere spanned
on the vector fields of velocity $u^{\alpha }$ and rotation $w^{\alpha } ($Ref.
1); II, in which only one Killing field is spanned on $u^{\alpha }$ and
$w^{\alpha } ($Ref. 2); and III, in which all Killing fields are linearly
independent of $u^{\alpha }$ and $w^{\alpha } ($Ref. 3). The many particular
cases arise because of several possible alignments or misalignments among the 3
Killing fields and $u^{\alpha }$ and $w^{\alpha }$.
\par
By the Bianchi type of the symmetry algebra and by the relation of the velocity
field to the symmetry orbits it can be recognized in which cases
generalizations of the Friedmann models can be expected. Two such candidate
cases were found in class II, and five more in class III. Those of class III
were prohibitively complicated, but one of the cases of class II allowed for
some progress, and this one is presented in the present paper. It is the
Bianchi type $V$ subcase of the case 1.2.2.2, given by eq. (5.19) in Ref. 2.
The other candidate case found in class II, eq. (5.10) in Ref. 2, can reproduce
only the de Sitter or the Einstein model in the limit of zero rotation, this is
seen from the time-dependence of the metric. Hence, it is not interesting for
cosmology and therefore disregarded here. In sec. 2, the metric is simplified
by a coordinate transformation, and a first integral of the Einstein equations
is found. With zero value of this integral, coordinate transformations can be
used to eliminate two components of the metric tensor, and the number of
nontrivial Einstein equations is reduced to 7. Although there are only 4
functions + matter density to be determined by these 7 equations, the set later
turns out to be self-consistent. In sec. 3, it is shown that the $k = -1$
Friedmann models are contained among the solutions that result in the limit of
zero rotation. In sec. 4, the Einstein equations are reduced to a set $S$ of 3
second-order equations to determine 3 metric components + a quadrature $Q$ to
determine the fourth component ($g_{33}$). Of the Einstein equations derived in
sec. 2, one is fulfilled identically in consequence of the set $\{S \bigcup
Q\}$, one turns out to be a constraint imposed on the initial data, and the one
that determines the matter-density turns out to provide a first integral. The
constraint and the first integral are second-degree polynomials in the first
derivatives of the unknown functions whose coefficients depend on the unknown
functions. The first integral determines $g_{33}$ algebraically in terms of the
other components, and so it is a replacement for the quadrature $Q$. It is also
shown that the set $S$ cannot be obtained as the Euler-Lagrange equations from
a variational principle of the Hilbert type. Finally, it is shown in sec. 4 how
the set $\{S \bigcup Q\}$ reproduces the Friedmann equations in the limit of
zero rotation and zero shear. In sec. 5, Lie point-symmetries of the set are
found: there is a two-dimensional symmetry group that allows one to reduce one
second-order equation to a first-order equation plus a quadrature. However,
this reduction provides no real progress toward solving the set $S$; the
first-order equation is still a member of a complicated set. In sec. 6, a
method of systematic search for polynomial first-order first integrals of a set
of ordinary differential equations is applied to the set $S$ of sec. 4. It is
shown that no first integrals that are polynomials of degree 1 or 2 in the
first derivatives exist. The same method is used to reveal the existence of a
possible constraint on initial data, which is of degree 1 in first derivatives,
that is preserved by the set $S$. However, the constraint necessarily implies
zero matter-density, and so it is not interesting for cosmology.

Calculations that are of secondary importance for the main text, but
are difficult to reproduce, are described in the appendices.

\bigskip

\section {The Einstein equations, their first integral and implications
of the zero value of this integral.}

\setcounter{equation}{0}

The subject of the present paper are the Einstein equations for the Bianchi
type V subcase of case 1.2.2.2 of Ref. 2. For reference, the initial formulae
are recalled in their original notation.

The Bianchi type V symmetry results when $c = 0$ in eqs. (5.19) of Ref. 2 and
when, in addition, $j = -a$ in eqs. (5.16). Hence, the metric is:

\begin{equation}
{\rm d}s^2 = {\rm d}t^2 + 2y{\rm d}t{\rm d}x +y^2h_{11}{\rm d}x^2 +
2h_{12}{\rm d}x{\rm d}y + 2y^2h_{13}{\rm d}x{\rm d}z
$$

$$
+ (h_{22}/y^2){\rm d}y^2 + 2h_{23}{\rm d}y{\rm d}z + y^2h_{33}{\rm d}z^2,
\end{equation}

\noindent where the coordinates are $\{x^{\alpha}\} = \{x^0, x^1, x^2, x^3\} =
\{t, x, y, z\}$, and $h_{ij}, i, j = 1, 2, 3$ are unknown functions of the
variable

\begin{equation}
v = {\rm e}^ty^{C_2/a},
\end{equation}

\noindent $a$ and $C_2$ being arbitrary constants. The velocity field
$u^{\alpha}$, the rotation field $w^{\alpha}$ and the Killing fields
${k_{(i)}}^{\alpha}$, $i = 1, 2, 3$ are given by:

\begin{equation}
u^{\alpha} = {\delta^{\alpha}}_0, \qquad w^{\alpha} =
(\rho/y){\delta^{\alpha}}_0, \qquad {k_{(1)}}^{\alpha} = {\delta^{\alpha}}_1,
\qquad {k_{(3)}}^{\alpha} = {\delta^{\alpha}}_3, $$

$$ {k_{(2)}}^{\alpha} = C_2{\delta^{\alpha}}_0 + a(x{\delta^{\alpha}}_1 -
y{\delta^{\alpha}}_2 + z{\delta^{\alpha}}_3),
\end{equation}

\noindent where $\rho$ is the matter-density of dust. The rotation tensor
$\omega_{\alpha \beta}$ has only one algebraically independent nonzero
component:

\begin{equation}
\omega_{12} = {1 \over 2},
\end{equation}

\noindent and therefore the coordinates used here are ill-suited for
considering the limit $\omega \to 0$. From the first equation in (2.3) it can
be seen that the coordinates are comoving.

As shown in Ref. 1, it follows from the equations of motion and from the
equation of conservation of the number of particles that:

\begin{equation}
g := {\rm det}(g_{\alpha \beta}) = - (y/\rho)^2,
\end{equation}

\noindent where $\rho$ is the mass-density.

This is the form in which the metric resulted from the Killing equations in
Ref. 2. It is advantageous to transform the coordinates as follows:

\begin{equation}
t = t' - (C_2/a)\ln y', \qquad x = x' - C_2/(ay'), \qquad (y, z) = (y', z').
\end{equation}

\noindent The result is equivalent to substituting $C_2 = 0$ and $a = 1$ in
eqs. (2.1) - (2.4), i.e. the forms of the metric (2.1), of the vector fields
$u^{\alpha}$, $w^{\alpha}$, ${k_{(1)}}^{\alpha}$ and ${k_{(3)}}^{\alpha}$ in
(2.3) and of the rotation tensor $\omega_{\alpha \beta}$ in (2.4) do not change
(although the new $h'_{ij}$ in (2.1) will be linear combinations of the old
$h_{ij}$), while the new ${k_{(2)}}^{\alpha}$ basis vector will be:

\begin{equation}
{k_{(2)}}^{\alpha} = x{\delta^{\alpha}}_1 - y{\delta^{\alpha}}_2 +
z{\delta^{\alpha}}_3,
\end{equation}

\noindent and the argument of $h_{ij}$ will now be $v = {\rm e}^{t'}$, i.e. the
$h_{ij}$ are from now on unknown functions of the time-coordinate $t$.

The isometry corresponding to (2.7) is:

\begin{equation}
t' = t, \qquad (x', z') = {\rm e}^{\tau}(x, z), \qquad y' = {\rm e}^{- \tau}y,
\end{equation}

\noindent where $\tau$ is the group parameter.

It is convenient to parametrize the metric as follows:

\begin{equation}
{\rm d}s^2 = ({\rm d}t + y {\rm d}x)^2 - (yK_{11}{\rm d}x)^2 - (K/y)^2({\rm d}y
+ y^2h{\rm d}x)^2 - {K_{33}}^2[yg{\rm d}x + (f/y){\rm d}y + y{\rm d}z]^2,
\end{equation}

\noindent where $K_{11}$, $K$, $K_{33}$, $h$, $f$, and $g$ are unknown
functions of $t$. The components of the Einstein tensor referred to below are
tetrad components $G_{IJ} = {e^{\alpha}}_I{e^{\beta}}_JG_{\alpha \beta}$, i.e.
projections of the coordinate components $G_{\alpha \beta}$ onto the
orthonormal tetrad $e^I := {e^I}_{\alpha}{\rm d}x^{\alpha}$ implied by (2.9):

\begin{equation}
e^0 = {\rm d}t + y{\rm d}x, \qquad e^1 = yK_{11}{\rm d}x, \qquad e^2 = (K / y)
({\rm d}y + y^2h{\rm d}x), $$

$$
e^3 = K_{33} [yg{\rm d}x + (f/y){\rm d}y + y{\rm d}z],
\end{equation}

\noindent where ${e^{\alpha}}_I$ is the inverse matrix to ${e^I}_{\alpha}$,
i.e. ${e^{\alpha}}_I{e^I}_{\beta} = {\delta^{\alpha}}_{\beta}$,
${e^{\alpha}}_J{e^I}_{\alpha} = {\delta^I}_J$. In the parametrization (2.9),
the determinant of the metric is:

\begin{equation}
g = - (y K_{11}KK_{33})^{2}.
\end{equation}

The tetrad components of the Einstein tensor corresponding to the metric (2.9)
are given in the Appendix A. As seen from there, two combinations of those
equations are of first order, they are $K_{11}G_{03} + G_{13} = 0$, i.e.:

\begin{equation}
({3 \over 2}K_{33} / K_{11}) [({K_{11}}^2 - 1) K^{-2}f,_{t} + h(hf,_{t} -
g,_{t})] = 0
\end{equation}

\noindent and ${K_{11}}{G_{02}} + {G_{12}} = 0$, i.e.:

\begin{equation}
({K_{11}}K)^{-1}[- {3 \over 2}K^2hh,_{t} + {1 \over 2}h - K_{11} {K_{11}},_{t}
+ ({K_{11}}^2 - 1) (2K,_{t} / K - {K_{33}},_{t} / K_{33})] = 0.
\end{equation}

\noindent As shown in Appendix B, the case $h = 0$ does not lead to interesting
developments, so we shall proceed further under the assumption:

\begin{equation}
h \neq 0.
\end{equation}

\noindent Then, eq. (2.12) implies:

\begin{equation}
g,_t = [h + ({K_{11}}^2 - 1)/(hK^2)]f,_t
\end{equation}

\noindent With this, the equations $G_{03} = G_{13} = G_{23} = 0$ turn out to
be equivalent, and they can be written as follows:

\begin{equation}
- {1 \over 2}\left({{{K_{11}}^2 - 1} \over h} \cdot {{{K_{33}}^3 f,_t} \over
{K_{11}K}}\right),_t + {{{K_{33}}^3 f,_t} \over {K_{11}K}} = 0.
\end{equation}

\noindent This invites the introduction of the new variable $u(t)$ by $u,_t =
h/({K_{11}}^2 - 1)$, and then (2.16) becomes:

\begin{equation}
\left({{{K_{33}}^3 f,_u} \over {K_{11}K}}\right),_u - 2{{{K_{33}}^3 f,_u} \over
{K_{11}K}} = 0,
\end{equation}

\noindent which has the first integral ${K_{33}}^3 f,_u /(K_{11}K) = C{\rm
e}^{2u}$, $C =$ const, i.e.:

\begin{equation}
f,_t = C{\rm e}^{2u}hK_{11}K/[{K_{33}}^3({K_{11}}^2 - 1)].
\end{equation}

\noindent From here on, we shall follow only the special case $C = 0$, which is
a solution of the Einstein equations, but not a general one: it is a subcase
chosen ad hoc for further progress with integration. Then, from (2.18) and
(2.15) $f =$ const, $g =$ const, and from (2.10) the coordinate transformation
$z' = z + f/y + gx$ leads to

\begin{equation}
f = g = 0
\end{equation}

\noindent without changing any of the other formulae for $g_{\alpha \beta}$,
$u^{\alpha}$, $w^{\alpha}$, $\omega_{\alpha \beta}$ or ${k_{(i)}}^{\alpha}$.

The Einstein equations $G_{03} = G_{13} = G_{23} = 0$ are now fulfilled
identically. We are left with 7 equations of the set (A.1) -- (A.10) in
Appendix A that should determine the 4 functions $K_{11}$, $K$, $K_{33}$ and
$h$, and the matter density $\rho$ in addition. It will turn out in sec. 4 that
the 7 equations are dependent just in the way needed to make the problem
self-consistent and determinate.

\section{The Friedmann limit of the metric.}

\setcounter{equation}{0}

As already stated, the coordinates used in sec. 2 are ill-suited for
considering the limit $\omega \to 0$. It will be shown in the present section
that this limit can be calculated after a coordinate transformation and a
reparametrization of the metric. This is just a demonstration of existence, and
it is not claimed that the limit $\omega \to 0$ thus obtained is unique (i.e.
another nonrotating limit might be obtained starting from a different
coordinate transformation). However, we will be satisfied to show that a limit
{\it exists} in which the $k < 0$ Friedmann model is contained.

Since $\omega_{12} = - \omega_{21} = {1 \over 2}$ are the only nonzero
components of the rotation tensor, a natural coordinate transformation to
consider is:

\begin {equation}
y = \omega_0 y'.
\end {equation}

\noindent where $\omega_0$ is a constant.
After the transformation:

\begin {equation}
\omega'_{12} = {1 \over 2}\omega_0 = - \omega'_{21}
\end {equation}

\noindent (all other $\omega_{\alpha \beta} = 0$), and the limit of zero
rotation is $\omega_0 \to 0$. However, before this limit is taken, the metric
functions in (2.9) must be reparametrized or else the limit will be singular.
The following reparametrizations will do the job:

\begin {equation}
K_{11} = \tilde{K}_{11} /\omega_0, \qquad K_{33} = \tilde{K}_{33} /\omega_0,
\qquad f = \tilde{f}\omega_0.
\end{equation}

\noindent The transformation (3.1) and the reparatmetrization (3.3) result in
the following metric:

\begin{equation}
{\rm d}s^2 = ({\rm d}t + \omega_0y'{\rm d}x)^2 - (y'\tilde{K}_{11}{\rm d}x)^2 -
K^2({\rm d}y'/y' + \omega_0y'h {\rm d}x)^2 $$ $$ - {\tilde{K}_{33}}^2[y'g {\rm
d}x + (\tilde{f}/y'){\rm d}y' + y'{\rm d}z]^2
\end{equation}

\noindent whose limit $\omega_0 \to 0$ (with primes and tildes omitted) is:

\begin{equation}
{\rm d}s^2 = {\rm d}t^2 - (yK_{11}{\rm d}x)^2 - (K/y)^2{\rm d}y^2 -
{K_{33}}^2[yg {\rm d}x + (f/y){\rm d}y + y{\rm d}z]^2,
\end{equation}

\noindent This is more than sufficiently general to accomodate the $k = -1$
Friedmann model that results when $g = f = 0$ and $K_{11} = K = K_{33} :=
R(t)$, where $R(t)$ is the Friedmann scale factor. The resulting coordinates
are none of the standard ones, but are related by $y = {\rm e}^u$ to one of the
sets used in the literature (see eq. (1.3.15) in Ref. 5).

The fact that (3.5), the limit $\omega_0 \to 0$ of (2.9), is still more general
than the Friedmann metric means that (3.5) has nonzero shear, i.e. shear
survives the transition $\omega \to 0$.

However, one possible problem still lies ahead. It was proven above that the $k
= -1$ Friedmann model is contained among the solutions of the set (A.1) --
(A.10). What is still needed is an explicit solution with the property that it
has nonzero rotation in general, but reproduces the $k = -1$ Friedmann model in
the limit $\omega \to 0$. Experience with the Einstein equations in other cases
shows that sometimes, while integrating the equations, one encounters mutually
exclusive alternatives $A$ and $B$ such that it is no longer possible to
recover $B$ as a limit of $A$ after the integration is completed. A well-known
example are the two subfamilies ($\beta' = 0$ and $\beta' \neq 0$) of the
Szekeres-Szafron$^{6-7}$ cosmological models; see Ref. 5 for more on this
point. (Only recently, a reformulation of the two classes was invented that
allows to recover the $\beta' = 0$ family from the other one, see Ref. 8).
Hence, it may still happen that among the explicit solutions, the rotating dust
model and the Friedmann $k = -1$ model will turn out to be mutually exclusive
subfamilies. This uncertainty will persist until an explicit solution is found.

It will be shown at the end of sec. 4 that the explicitly written out Einstein
equations do allow a continuous limiting transition $\omega \to 0, \sigma \to
0$, and in the limit they reproduce exactly the Friedmann equations.

\section{The independent Einstein equations.}

\setcounter{equation}{0}

We shall now proceed with the subcase (2.19). Eq. (2.12) is then fulfilled
identically. Eq. (2.13) does not change, and it can be more conveniently
rewritten if $K_{11}$ is parametrized as follows:

\begin{equation}
K_{11} = \cosh(F).
\end{equation}

\noindent Then, from (2.13):

\begin{equation}
{K_{33}},_t = K_{33}[- {3 \over 2}K^2hh,_t / \sinh^2(F) + {1 \over 2}h /
\sinh^2(F) + 2K,_t/K - \cosh(F)F,_t / \sinh(F)].
\end{equation}

\noindent When this is substituted into the remaining equations (A.1) --
(A.10), the function $K_{33}$ disappears from the set completely, i.e. we are
left with 6 equations to determine $h$, $K$, $F$ and the matter-density plus
the quadrature implied by (4.2) that allows one to calculate $K_{33}$ once
$h(t)$, $K(t)$ and $F(t)$ are known.

Since (2.13) is now satisfied, the equations $G_{02} = 0$ and $G_{12} = 0$ are
equivalent, and they can be written as:

\begin{equation}
h,_{t t} = {3 \over 2}K^{2}h{h,_{t}}^{2} /{\sinh}^{2}(F) - 5K,_{t}h,_{t}/K +
(2{\cosh}^{2}(F) - 1)F,_{t}h,_{t}/{\sinh}(F)\cosh(F) $$

$$ + hh,_{t}/{\sinh}^{2}(F) + K,_{t}/K^{3} + F,_{t} /K^{2}{\cosh}(F){\sinh}(F)
- {1 \over 2}h/(K{\sinh}(F)) ^{2}
\end{equation}

\noindent This is used to eliminate $h,_{tt}$ from the other Einstein
equations. The equation $G_{01} = 0$ can then be solved for $F,_{tt}$ (the
solution is given in Appendix C) and this is used to eliminate $F,_{tt}$ from
the diagonal components of the Einstein tensor (all the non-diagonal Einstein
equations have been used up at this point). After such a substitution, the
following identity is fulfilled:

\begin{equation}
G_{11} + G_{33} - 2G_{22} \equiv 0,
\end{equation}

\noindent i.e. one of the three equations $G_{11} = G_{22} = G_{33} = \Lambda$
can be discarded because it is a consequence of the remaining two. We choose to
discard $G_{33} = \Lambda$.

Then, $K,_{tt}$ can be calculated from $G_{22} - G_{11} = 0$. The result is:

\begin{equation}
K,_{t t} = {1 \over 4}K^{3}{\sinh}^{-2}(F){h,_{t}}^{2} - {3 \over 2}K^{3}h\cosh
(F){\sinh}^{-3}(F)F,_{t}h,_{t} $$

$$ - {\cosh}^{-1}(F){\sinh}^{-1}(F)K,_{t}F,_{t} + 2\cosh
(F){\sinh}^{-1}(F)K,_{t}F,_{t} - K{\cosh}^{2}(F){\sinh}^{-2}(F){F,_{t}}^{2} $$

$$ - {3 \over 4}Kh,_{t} + {3 \over 2}K^{3}h^{2}{\sinh}^{-4}(F)h,_{t} + {3 \over
4}K^{3}h^{2}{\sinh}^{-2}(F)h,_{t} $$

$$ - {3 \over 2}hK,_{t} - h{\sinh}^{-2}(F)K,_{t} -
Kh{\cosh}^{-1}(F){\sinh}^{-1}(F)F,_{t} + {3 \over
2}Kh{\cosh}^{3}(F){\sinh}^{-3}(F)F,_{t} $$

$$ - {1 \over 4}K^{-1}{\cosh}^{2}(F){\sinh}^{-2}(F) - {1 \over
2}Kh^{2}{\sinh}^{-4}(F) - {1 \over 4}Kh^{2}{\sinh}^{-2}(F)
\end{equation}

\noindent This is used to eliminate $K,_{tt}$ from the right-hand side of the
equation determinig $F,_{tt}$ (see Appendix C), and the result is:

\begin{equation}
F,_{t t} = - {3 \over 4}K^{2}{\cosh}^{-1}(F){\sinh}^{-1}(F){h,_{t}}^{2} - {3
\over 2}Kh{\cosh}^{-1}(F){\sinh}^{-1}(F)K,_{t}h,_{t} $$

$$ + 2K^{-2}{\cosh}^{-1}(F)\sinh (F){K,_{t}}^{2}- K^{-1}K,_{t}F,_{t} - \cosh
(F){\sinh}^{-1}(F){F,_{t}}^{2} $$

$$ + {3 \over 4}K^{2}h^{2}{\cosh}^{-1}(F){\sinh}^{-1}(F)h,_{t} + {3 \over
2}K^{2}h^{2}{\cosh}^{-1}(F){\sinh}^{-3}(F)h,_{t} $$

$$ + {1 \over 2}{\cosh}^{-1}(F){\sinh}^{-1}(F)h,_{t} - {3 \over
4}{\cosh}^{-1}(F)\sinh (F)h,_{t} $$

$$ - {5 \over 2}K^{-1}h{\cosh}^{-1}(F){\sinh}^{-1}(F)K,_{t} - {3 \over
2}K^{-1}h{\cosh}^{-1}(F)\sinh (F)K,_{t} $$

$$ + h{\sinh}^{-2}(F)F,_{t} + {3 \over 2}hF,_{t} - {1 \over
4}K^{-2}{\cosh}^{-1}(F)\sinh (F) - {3 \over
4}K^{-2}{\cosh}^{-1}(F){\sinh}^{-1}(F) $$

$$
- {1 \over 2}h^{2}{\cosh}^{-1}(F){\sinh}^{-3}(F)
- {1 \over 4}h^{2}{\cosh}^{-1}(F){\sinh}^{-1}(F)
\end{equation}

\noindent With (4.2), (4.3), (4.5) and (4.6) all substituted into (A.5), the
equation $G_{11} = \Lambda$ reduces to the following form:

\begin{equation}
G_{11} = {1 \over 4}K^{2}{\cosh}^{-2}(F){h,_{t}}^{2} + {3 \over
2}Kh{\cosh}^{-2}(F)K,_{t}h,_{t} + {3 \over
2}K^{2}h{\cosh}^{-1}(F){\sinh}^{-1}(F)F,_{t}h,_{t} $$

$$ - 2K^{-2}{\cosh}^{-2}(F){\sinh}^{2}(F){K,_{t}}^{2} -
2K^{-1}{\cosh}^{-1}(F)\sinh (F)F,_{t}K,_{t} + {F,_{t}}^{2} $$

$$ + {3 \over 2}K^{2}h^{2}{\cosh}^{-2}(F)h,_{t} -
3K^{2}h^{2}{\sinh}^{-2}(F)h,_{t} - {1 \over 2}{\cosh}^{-2}(F)h,_{t} + {3 \over
2}h,_{t} $$

$$ + {5 \over 2}K^{-1}h{\cosh}^{-2}(F)K,_{t} + 3K^{-1}hK,_{t} - {5 \over
2}h{\cosh}^{-1}(F){\sinh}^{-1}(F)F,_{t} - 3h{\cosh}^{-1}(F)\sinh (F)F,_{t} $$

$$ + {1 \over 4}K^{-2}{\cosh}^{-2}(F) + {3 \over 2}K^{-2} + {1 \over
2}h^{2}{\cosh}^{-2}(F) + h^{2}{\sinh}^{-2}(F) = \Lambda.
\end{equation}

\noindent Now it may be verified that $G_{11} =$ const is preserved by eqs.
(4.3), (4.5) and (4.6). This is done as follows. The derivative ${{\rm d} \over
{{\rm d}t}}G_{11}$ is calculated, and $h,_{tt}$, $K,_{tt}$ and $F_{tt}$ that
reappear are eliminated using (4.3), (4.5) and (4.6). Then, ${K,_t}^2$ is found
from (4.7) and used to eliminate ${K,_t}^3$ and ${K,_t}^2$ from ${{\rm d} \over
{{\rm d}t}}G_{11}$. The result is the identity ${{\rm d} \over {{\rm
d}t}}G_{11} \equiv 0$. This means that, in virtue of the other field equations,
if $G_{11} = \Lambda$ holds at any given time, then it will remain constant at
all other times. Hence, $G_{11} = \Lambda$ is a limitation imposed by the
Einstein equations on the initial data for eqs. \{(4.3), (4.5), (4.6)\}, and it
defines the cosmological constant in terms of the other constants that will
appear after (4.3), (4.5) and (4.6) are solved. If $\Lambda = 0$, then $G_{11}
= 0$ reduces the number of arbitrary constants by 1.

Hence, with (4.4), we are left with only four equations: (4.2), (4.7) and any
two equations from the set $S = \{(4.3), (4.5), (4.6)\}$, to determine the four
functions $K_{33}$, $h$, $K$ and $F$. The third equation in $S$ is implied by
the remaining two together with (4.7). The only field equation that has not yet
been used up is:

\begin{equation}
G_{00} = (8\pi G/c^4)\rho - \Lambda.
\end{equation}

\noindent This may be expected to simply define the matter-density in terms of
the metric functions. However, in the formulation used in this paper,
matter-density enters the equations in two ways: as a source term in $G_{00}$
above, and also through (2.5). From (2.5) and (2.11) it follows that $\rho$
must be related to the other functions by:

\begin{equation}
\rho = (K_{11}KK_{33})^{-1}.
\end{equation}

\noindent Together with (4.8) and (4.1) this implies that the following must hold:

\begin{equation}
[(G_{00} + \Lambda)\cosh(F)KK_{33}],_t \equiv 0.
\end{equation}

\noindent Indeed, this is an identity. This is verified as follows. First,
(4.1), (4.2), (4.3), (4.5) and (4.6) are substituted into (A.1) (with $f = g =
0$) to eliminate all second derivatives. Then, (4.10) is calculated, and (4.2),
(4.3), (4.5) and (4.6) are used to eliminate ${K_{33}},_t$ and all second
derivatives again. Finally, (4.7) is used to eliminate ${K,_t}^3$ and
${K,_t}^2$ from the left-hand side of (4.10). In the end, the identity (4.10)
results. Hence, (2.5) and (4.8) are consistent in virtue of the other field
equations, and moreover $(G_{00} + \Lambda)\cosh(F)KK_{33} = C = {\rm const}$
(with second derivatives of $h$, $K$ and $F$ eliminated by (4.3), (4.5) and
(4.6) and with ${K,_t}^2$ eliminated by (4.7)) is the following first integral
of the Einstein equations:

\begin{equation}
K_{33}[- 3Kh\sinh (F)F,_{t} - {3 \over 2}Kh^{2}{\cosh}^{-1}(F) + {3 \over
2}K{\cosh}^{-1}(F){\sinh}^2(F)h,_{t} $$

$$+ 3h{\cosh}^{-1}(F){\sinh}^2(F)K,_{t} - {3 \over
2}K^{-1}{\sinh}^2(F){\cosh}^{-1}(F) - {3 \over
2}K^{3}h^{2}{\cosh}^{-1}(F)h,_{t}] = C.
\end{equation}

\noindent Note that, from (4.8) and (4.9), $C = 8\pi G/c^4 \neq 0$, and so
(4.11) determines $K_{33}$ algebraically. Hence, (4.11) can replace (4.2) as
the definition of $K_{33}$. Thereby, the problem of this paper was reduced to
the following procedure:

1. Find the most general solution of the set $\{(4.3), (4.5), (4.6)\}$. It will
contain 6 arbitrary constants $\{C_1, \dots , C_6\}$.

2. Impose (4.7) on the $\{h, K, F\}$ found in the previous step. This will be
just a definition of $\Lambda$ in terms of $\{C_1, \dots , C_6\}$ or, when
$\Lambda = 0$, an additional constraint imposed on $\{C_1, \dots , C_6\}$.

3. Calculate $K_{33}$ from (4.11), with $C = 8\pi G/c^4$.

4. Calculate the matter-density from (4.9).

As shown in Ref. 9, an efficient method to find first integrals of a set of
equations exists if the set can be obtained from a Lagrangian. Unfortunately,
the problem of determinig whether a given set of equations is derivable from a
lagrangian is rather complicated and unsolved in general $^{10}$. It is known
that the Einstein equations for class B Bianchi metrics may not admit a
lagrangian, even though the general Einstein equations do (see Ref. 11 for an
explanation). It is shown in Appendix D that eqs. $\{4.3), (4.5), (4.6)\}$ do
not follow from the most natural lagrangian conceivable in this case: a
second-degree polynomial in the first derivatives of $h$, $K$ and $F$, with
coefficients being functions of $h$, $K$ and $F$.

For further reference, let us consider the limit of zero rotation in (4.2) --
(4.3) and (4.5) -- (4.7). After the reparametrization (3.3) we have:

\begin{equation}
\cosh (F) = \tilde{{K}_{11}} / \omega_0, \qquad \sinh (F) = \sqrt {{\left.
{\tilde{K}}_{11} \right. }^2 / {\omega_0}^2 - 1}, $$

$$ F,_t = \tilde{{K}_{11,t}} / \sqrt {{\left. {\tilde{K}}_{11} \right. }^2 -
{\omega_0}^2} ,
\end{equation}
and then (4.2) in the limit $\omega_0 \to 0$ becomes:

\begin{equation}
\tilde{K}_{33,t} = \tilde{K}_{33}(2K,_t/K - \tilde{{K}_{11,t}}/
\tilde{{K}_{11}}),
\end{equation}
which is an identity in the Friedmann limit $\tilde{{K}_{11}} = K =
\tilde{K}_{33} = R(t)$.

Eq. (4.3) could in fact be discarded in the limit $\omega_0 \to 0$. This is
because eq. (4.3) was derived from (A.3), and those terms in (A.3) that lead to
(4.3) are all multiplied by ${\omega_0}^2$ after the reparametrization (3.3).
The off-diagonal component of (3.4) that is proportional to $h$ will vanish
with any $h$ when $\omega_0 \to 0$. Nevertheless, (4.3) gives a result
consistent with the other equations in this limit. The limiting form of it is:

\begin{equation}
h,_{tt} = - 5h,_tK,_t/K + 2 \tilde{{K}_{11,t}}h,_t/ \tilde{{K}_{11}} +
K,_t/K^3.
\end{equation}

The limit $\omega_0 \to 0$ of (4.5) is:
\begin{equation}
K,_{tt} = - K{\left. {\tilde{K}}_{11,t} \right. }^2/{\left. {\tilde{K}}_{11}
\right. }^2 - {3 \over 4}Kh,_t - {3 \over 2}hK,_t + {3 \over 2}Kh
\tilde{{K}_{11,t}}/ \tilde{{K}_{11}} - 1 / (4K) + 2K,_t \tilde{{K}_{11,t}}/
\tilde{{K}_{11}}.
\end{equation}
The same limit of (4.6) is:

\begin{equation}
\tilde{K}_{11,tt}/ \tilde{{K}_{11}} = 2{K,_t}^2/K^2 - K,_t \tilde{{K}_{11,t}}/
(K\tilde{{K}_{11}}) + {3 \over 2}h \tilde{{K}_{11,t}}/ \tilde{{K}_{11}} - {3
\over 2}h,_t - {3 \over 2}hK,_t/K - 1/(4K^2).
\end{equation}

In the Friedmann limit $\tilde{{K}_{11}} = K = R(t)$, eqs. (4.15) and (4.16)
become identical:

\begin{equation}
R,_{tt}/R = {R,_t}^2/R^2 - {3 \over 4} h,_t - 1 / (4R^2).
\end{equation}
Finally, the limit $\omega_0 \to 0$ of (4.7) is:

\begin{equation}
- 2 {K,_t}^2/K^2 - 2K,_t \tilde{{K}_{11,t}}/ (K \tilde{{K}_{11}}) + {\left.
{\tilde{K}}_{11,t} \right. }^2/{\left. {\tilde{K}}_{11} \right. }^2 + {3 \over
2}h,_t + 3hK,_t/K - 3h \tilde{{K}_{11,t}}/ \tilde{{K}_{11}}$$

$$ + 3 / (2K^2) = \Lambda.
\end{equation}
The Friedmann limit of this is:

\begin{equation}
- 3 {R,_t}^2/R^2 + {3 \over 2}h,_t + 3 / (2R^2) = \Lambda.
\end{equation}
Finding $h,_t$ from (4.19) and substituting it in (4.17) we obtain:

\begin{equation}
R,_{tt}/R = - {R,_t}^2 / (2R^2) + 1 / (2R^2) - \Lambda /2,
\end{equation}
which is exactly one of the Friedmann equations. Incidentally, the $h,_t$ found
from (4.19), if substituted in (4.14), leads to (4.20) again. Hence, in the
Friedmann limit, (4.14) follows from (4.19) and (4.17), and need not be
discarded.

Note that also (4.11) has a meaningful Friedmann limit. In order to make this
limit finite, it must be assumed that:

\begin{equation}
C = \tilde{C}/{\omega_0}^2,
\end{equation}
and then the limit $\omega_0 \to 0$ of (4.11) is:

\begin{equation}
K_{33}[\Lambda K \tilde{{K}_{11}} + 2{K,_t}^2 \tilde{{K}_{11}}/K + 2K,_t
\tilde{{K}_{11,t}} - K{\left. {\tilde{K}}_{11,t} \right. }^2/ \tilde{{K}_{11}}
- 3 \tilde{{K}_{11}}/K] = \tilde{C}.
\end{equation}
In the Friedmann limit this becomes:

\begin{equation}
R(\Lambda R^2 + 3{R,_t}^2 - 3) = \tilde{C}.
\end{equation}
Recalling the Friedmann formula for the mass-density, with $k = -1$:

\begin{equation}
3{R,_t}^2/R^2 - 3/R^2 + \Lambda = (8\pi G/c^2)\rho,
\end{equation}
we recognize in (4.23) the familiar mass-conservation formula of the Friedmann
model, $\rho R^3 = c^2\tilde{C}/(8\pi G) = {\rm const}$.

\section {The Lie point-symmetries of the equations (4.3), (4.5) and (4.6).}

\setcounter{equation}{0}

Point symmetries of (sets of) differential equations are transformations in the
space of the independent + dependent variables that leave the set of solutions
of the equations unchanged. The point symmetries that form Lie groups (if they
exist for a given set of equations) can help in transforming apparently
intractable equations into solvable ones by adapting the variables suitably to
the generators of the symmetries. The background philosophy and many of the
methods are analogous to simplifying the Einstein equations by adapting the
coordinates to the Killing vector fields (if such exist). It is assumed that
the readers are familiar with this latter procedure. The basic definitions and
theorems concerning point symmetries are presented in detail in Refs. 9 and 10.

Eqs. (4.3), (4.5) and (4.6) are of the following form:

\begin{equation}
{{{\rm d}^2z^i} \over {{\rm d}t^2}} = {W^i}_{jk}{{{\rm d}z^j} \over {{\rm d}t}}
{{{\rm d}z^k} \over {{\rm d}t}} + {V^i}_j{{{\rm d}z^j} \over {{\rm d}t}} + U^i,
\end{equation}

\noindent where $i = 0, 1, 2$; $(z^0, z^1, z^2) = (h, K, F)$ and ${W^i}_{jk}$,
${V^i}_j$ and $U^i$ are functions of the $z^i$, but not of $t$. (Incidentally,
the independence of $t$ of all these coefficients immediately implies one group
of symmetries, $t \to t' = t + s$, where $s$ is the group parameter. This group
will emerge from the calculation below.) Let the following be a one-dimensional
group of point transformations:

\begin{equation}
t' = t'(t,\{z^j\}, \tau), \qquad z'^i = z'^i(t,\{z^j\}, \tau),
\end{equation}

\noindent where $\tau$ is the group parameter and $\tau = \tau_0$ corresponds
to the identity (so that $t'(t,\{z^j\}, \tau_0) \equiv t$, etc.). The
generators of this group (the field of vectors tangent to the orbits of the
group (5.2)) are then:

\begin{equation}
X = \xi {{\partial} \over {\partial t} } + \eta^j {{\partial} \over {\partial
{z^j}} },
\end{equation}

\noindent where:

\begin{equation}
\left[ \begin{array}{c} \xi \\ \eta^j \end{array} \right] = {{\rm d} \over
{{\rm d} \tau}} \left[ \begin{array}{c} t' \\ z'^j \end{array} \right]_{\tau =
\tau_0} .
\end{equation}

\noindent The generator $X$ is extended to arbitrary derivatives ${{{\rm d}^kz}
\over {{\rm d}t^k}} := {\stackrel {(k)}z}$ by the recursive formulae:

\begin{equation}
{\stackrel {(0)j} {\eta}} = \eta^j, \qquad {\stackrel {(k)j} {\eta}} = {{{\rm
d}{\stackrel{(k-1)j}{\eta}}} \over {{\rm d}t}} - {{{\rm d}^kz^j} \over {{\rm
d}t^k}}{{{\rm d}\xi} \over {{\rm d}t}},
\end{equation}

\noindent and by:

\begin{equation}
{\stackrel{(k)}X} = \xi {{\partial} \over {\partial t} } + \eta^j{{\partial}
\over {\partial z^j} } + {\stackrel {(1)j} {\eta}} {{\partial} \over {\partial
{\stackrel{(1)j}{z}}}} + \dots + {\stackrel {(k)j} {\eta}} {{\partial} \over
{\partial {\stackrel{(k)j}{z}}}} .
\end{equation}

\noindent The derivatives ${{\rm d} \over {{\rm d}t}}$ in (5.5) are total
derivatives, i.e.

$$ {{\rm d} \over {{\rm d}t}}f(t,\{z^i\}, \{{\stackrel {(1)i} z}\}, \dots ,
\{{\stackrel {(k)i} z}\}) = {{\partial f} \over {\partial t}} + {{{\rm d}z^j}
\over {{\rm d}t}} {{\partial f} \over {\partial z^j} } + \sum_{p=1}^k
{\stackrel {(p+1)j} z} {{\partial} \over {\partial {\stackrel{(p)j}{z}}}} , $$

\noindent and the order $n$ to which the generator $X$ has to be extended is
equal to the highest order of derivatives in the set (5.1) ($n = 2$ in our
case). A generator of a point symmetry obeys then:

\begin{equation}
{\stackrel{(n-1)} X}\Omega^i = {{{\rm d} {\stackrel{(n-1)i} {\eta}}} \over
{{\rm d}t}} - \Omega^i {{{\rm d}\xi} \over {{\rm d}t}},
\end{equation}

\noindent where $\Omega^i$ is the right-hand side of (5.1). (The right-hand
side of (5.7) is the ${\stackrel{(n)i} {\eta}}$ as given by (5.5), but with
${{{\rm d}^n z^i} \over {{\rm d}t^n}}$ replaced by $\Omega^i$ from (5.1)). Eqs.
(5.7) must be identities in all the derivatives ${\stackrel{(1)i} {z}}, \dots ,
{\stackrel{(n-1)i} {z}}$, and so they imply several separate equations to be
obeyed by the $\xi$ and $\eta^i$.

The procedure in finding and exploiting point symmetries is thus the following:

1. Find the general solution of (5.7) for $X$. Since the generators form a Lie
algebra (see Ref. 9), the most general $X$ will be spanned on a finite number
of basis vector fields $X_{(k)}$.

2. Read off the basis $X_{(k)}$ from that solution.

3. Adapt the variables $\{t'(t,\{z^j\}), z'^i(t, \{z^j\})\}$ to the basis
fields $X_{(k)}$ so as to maximally simplify the equations.

For our equations (5.1), eqs. (5.7) imply the following four relations:

\begin{equation}
\xi,_{kl} + {W^j}_{kl}\xi,_j = 0,
\end{equation}

\bigskip

\begin{equation}
{\eta^i},_{kl} = {W^i}_{kl,s}\eta^s + 2{W^i}_{s(l}{\eta^s},_{k)} -
{W^s}_{kl}{\eta^i},_s + {\delta^i}_{(l}{V^s}_{k)}\xi,_s + {V^i},_{(l}\xi,_{k)}
+ 2 {{\partial^2 \xi} \over {{\partial t}{\partial z^{(k}}}}{\delta^i}_{l)} ,
\end{equation}

\noindent where parentheses on indices denote symmetrization,

\begin{equation}
{{\partial^2 \eta^i} \over {{\partial t}{\partial z^k}}} = {W^i}_{ks}
{{\partial \eta^s} \over {\partial t} } + {1 \over 2} {V^i}_{k,s}\eta^s + {1
\over 2} {V^i}_s{\eta^s},_k - {1 \over 2} {V^s}_k{\eta^i},_s $$

$$ + {1 \over 2} {V^i}_k {{\partial \xi} \over {\partial t} } + U^i\xi,_k + {1
\over 2} {\delta^i}_k U^s\xi,_s + {1 \over 2} {\delta^i}_k {{\partial^2 \xi}
\over {\partial t^2} },
\end{equation}

\bigskip
\begin{equation}
{{\partial^2 \eta^i} \over {\partial t^2} } = {V^i}_s {{\partial \eta^s} \over
{\partial t} } + {U^i},_s\eta^s - U^s{\eta^i},_s + 2U^i {{\partial \xi} \over
{\partial t} } .
\end{equation}

\noindent The general solution of these equations (with ${W^i}_{kl}$, ${V^i}_k$
and $U^i$ read off from (4.3), (4.5) and (4.6)) is:

\begin{equation}
X = A {{\partial} \over {\partial t} } + B (t{{\partial} \over {\partial t} } -
h {{\partial} \over {\partial h} } + K {{\partial} \over {\partial } K}) ,
\end{equation}

\noindent where $A$ and $B$ are arbitrary constants. The proof that this is the
most general solution is laborious but straightforward, it is given in Appendix
E. Hence, our set of equations has a two-dimensional symmetry group whose
generators are:

\begin{equation}
X_{(1)} = {{\partial} \over {\partial t} }, \qquad X_{(2)} = t{{\partial} \over
{\partial t} } - h {{\partial} \over {\partial h} } + K {{\partial} \over
{\partial } K} ,
\end{equation}

\noindent and the corresponding finite symmetry transformations are:

\begin{equation}
t' = t + \tau_1, \qquad (h', K', F') = (h, K, F);
$$

$$ t' = {\rm e}^{\tau_2}t, \qquad h' = {\rm e}^{- \tau_2}h, \qquad K' = {\rm
e}^{\tau_2}K, \qquad F' = F,
\end{equation}

\noindent where $\tau_1$ and $\tau_2$ are the group parameters. The first
symmetry was self-evident, as already mentioned, and the second one can be
verified by inspection of the equations (4.3), (4.5) and (4.6).

Unfortunately, these symmetries do not lead to any discernible simplification
of the set $S = \{(4.3), (4.5), (4.6)\}$. In variables adapted to the generator
$X_{(1)}$, the independent variable is $K$, and $t(K)$ is one of the functions.
The set (5.1) thus transformed is of first order in $\phi(K) := dt/dK$, but the
first-order equation is still a member of a complicated set and none of the
equations separates out. Moreover, after the transformed set is algebraically
solved for $t,_{KK}$, $h,_{KK}$ and $F,_{KK}$, the right-hand sides become
polynomials of {\it third} degree in $t,_k$, $h,_K$ and $F,_K$.

The variables adapted to the generator $X_{(2)}$ are $(t', h', K')$, where:

\begin{equation}
t = {\rm e}^{K'}t', \qquad K = {\rm e}^{K'}, \qquad h = {\rm e}^{- K'}h'.
\end{equation}

\noindent In these variables, the set (5.1) becomes of first order in $\psi(t')
= K',_{t'}$. However, after it is solved for $h',_{t't'}$, $K',_{t't'}$ and
$F,_{t't'}$, the right-hand sides of $h',_{t't'}$ and $F,_{t't'}$ contain
rational functions of the form $W/(1 + t'K',_{t'})$, where $W$ is a monomial of
second degree in some of the $h',_{t'}$, $K',_{t'}$ and $F,_{t'}$. Neither
equation separates out. It is not possible to adapt the variables to both the
generators simultaneously because the group is nonabelian. This author was not
able to make any use of the new variables.

\section {First integrals that are polynomials in $(h,_t, K,_t, F,_t)$.}

\setcounter{equation}{0}

Suppose that the set $\tilde{S} = \{(4.3), (4.5), (4.6), (4.7)\}$ has a first
integral of the form:

\begin{equation}
I := Q_{ij}\dot{z}^i\dot{z}^j + L_i\dot{z}^i + E = C = {\rm const},
\end{equation}
where $C$ is an arbitrary constant, $Q_{ij} = Q_{ji}$, $L_i$ and $E$ are
unknown functions of $(h, K, F)$, $i, j = 1, 2, 3$, $z^1 = h, z^2 = K, z^3 =
F$. Then ${{{\rm d}I} \over {{\rm d}t}} \equiv 0$ in virtue of $\tilde{S}$,
i.e. using (5.1) to eliminate $\ddot{z}^i$:

\begin{equation}
(2Q_{ij}\dot{z}^j + L_i)({W^i}_{kl}\dot{z}^k\dot{z}^l + {V^i}_k\dot{z}^k + U^i)
+ Q_{ij,k}\dot{z}^i\dot{z}^j\dot{z}^k + L_{i,j}\dot{z}^i\dot{z}^j +
E,_i\dot{z}^i = 0.
\end{equation}
In showing that (6.2) is zero, (4.7) must be used. Eq. (4.7) may be safely used
to eliminate ${F,_t}^3$ and  ${F,_t}^2$, but not the remaining $F,_t$. This is
because $F,_t$ found from (4.7) would be of the form:

\begin{equation}
F,_t = P(h,_t , K,_t) + \sqrt { \Delta (h,_t , K,_t)},
\end{equation}
where $P$ and $\Delta$ are polynomials of degree 2 in $h,_t$ and $K,_t$. If
$\Delta$ were a square of a first-degree polynomial, then (6.3) could be used
to eliminate $F,_t$ altogether from (6.2). However, $\Delta$ being a square
implies an additional equation obeyed by $h,_t$ and $K,_t$ (the discriminant of
$\Delta$ must be zero). Hence, if $h,_t$ and $K,_t$ are to be treated as
independent, then $F,_t$ is linearly independent of $h,_t$ and $K,_t$. Then the
coefficients of $F,_t$ in (6.2) must sum up to zero anyway, and eliminating
$F,_t$ is of no use.

Knowing this, it can be verified that first integrals of the form (6.1) do not
exist. The calculations are conceptually straightforward, but lead through
horrible intermediate expressions, so they are not reported here. The
hypothesis that (6.1) is a first integral uniquely leads to an equation that is
equivalent to (4.7).

The same method may be used to test whether our set of equations admits a
constraint that would be a polynomial of degree 1 or 2 in the first
derivatives. The only difference with respect to the procedure of looking for a
first integral is that in verifying whether (6.2) is zero, eq. (6.1) is used,
too. If a nontrivial solution of (6.2) with this additional simplification is
found, then it means that the derivative of (6.1) by $t$ is zero if (6.1) holds
for any fixed $t$. Then, such (6.1) is a constraint preserved by the set $S$.
However, even this attempt has not led to useful results. Constraints of degree
2, i.e. those with $Q_{ij} \neq 0$, lead to prohibitively complicated equations
and could not be investigated. One constraint of the form (6.1) with $Q_{ij} =
0$ was found, but it is equivalent to the square bracket in (4.11) being zero,
and so implies zero matter density. Again, the details are not reported because
they contain complicated equations, but no ingenious ideas. This result proves
the usefulness of the method -- a sensible constraint was revealed -- but the
solution with zero density is not interesting for cosmology, and thus not
necessarily worth investigating.

The zero-density constraint was found without using eq. (4.7). Eq. (4.7) would
reduce the number of unknown functions by one, but the resulting set of
equations is prohibitively complicated and no progress was achieved.

\section {Summary of results.}

\setcounter{equation}{0}

It was shown that the Einstein equations for the metric (2.9) with $f = g = 0$
are self-consistent and solvable. They reduce to the set $S = \{(4.3), (4.5),
(4.6)\}$ to determine $h$, $K$ and $K_{11} = \cosh(F)$, and (4.11) to determine
$K_{33}$ (where $C = 8\pi G/c^4$). The matter density is found from (4.9). The
first derivatives of the functions obeying the set $S$ must obey (4.7).

The Friedmann solution with $k = -1$ is contained among the solutions of this
set, as shown in eqs. (4.12) -- (4.24). Unfortunately, no explicit example of a
more general solution could be found. Attempts to follow ad hoc Ansatzes
produced uninteresting results. The Ansatz $K = K_{33}$ led to the deSitter
solution in disguise, in which the t-lines had nonzero rotation. The Ansatz
$K_{11} = K / C$ ($C =$ const), which is consistent with the Friedmann limit,
led to such complicated equations that it could not even be verified if they
are not contradictory. The assumption of zero shear implies zero expansion, in
virtue of the theorem $(\sigma = 0) \Rightarrow (\omega \theta = 0)$ that holds
for dust (see Ref. 12).

The set $S$ was shown to have a two-dimensional group of point symmetries,
given by (5.14), and to admit no Lagrangian of the Hilbert type. It was also
verified that no first integrals of the form (6.1) exist.

The progress achieved in this paper was the reduction of the problem of
existence of a rotating generalization of the $k = -1$ Friedmann model to the
technical problem of finding an explicit solution of the set $S$. The
solvability of the set $S$ may be taken for granted because the Friedmann model
itself was shown to be one of its solutions. It is still unknown, though,
whether a continuous family of solutions exists labeled by by the parameter
$\omega$ (rotation) such that the limit $\omega \to 0$ taken in the explicit
solution leads to the $k = -1$ Friedmann model.

A similar analysis as done here should be done for the other promising cases
identified in Ref. 3.

\bigskip

{\bf Acknowledgements.} The algebraic manipulations for this paper were carried
out using the computer algebra system Ortocartan$^{13, 14}$ that was for this
purpose extended by several new programs$^{15}$. The author is grateful to J.
Kijowski, J. Jezierski, K. Rosquist, R. Jantzen, M. A. H. MacCallum and C.
Uggla for useful comments and instructions on the Lagrangian methods.

\begin{center}

{\bf Appendix A}

{\bf The Einstein equations for the metric (2.9).}

\end {center}

As explained in sec. 2 (after eq. (2.9)), these are the projections of the
Einstein tensor on the forms of the orthonormal tetrad (2.10), thus for example
$G_{03}$ below is equal to ${e_I}^{\alpha} {e_J}^{\beta} G_{\alpha \beta}$ with
I = 0 and J = 3, where $G_{\alpha \beta}$ are the coordinate components of the
Einstein tensor.

$$ G_{00} = 2{K_{11}}^{-3}h{K_{11}},_{t} + {K_{11}}
^{-3}K^{-1}{K_{11}},_{t}K,_{t} +
{K_{11}}^{-3}{K_{33}}^{-1}{K_{11}},_{t}{K_{33}},_{t} $$

$$ + {3 \over 4}{K_{11}}^{-2}K^{-2} - {1 \over
4}{K_{11}}^{-2}K^{-2}{K_{33}}^{2}{f,_{t}}^{2} - 3{K_{11}}^{-2}K^{-1}hK,_{t} $$

$$ - {K_{11}}^{-2}K^{-1}{K_{33}}^{-1}K,_{t}{K_{33}},_{t} -
{K_{11}}^{-2}K^{-1}K,_{t t} - {1 \over 4}{K_{11}}^{-2}K^{2}{h,_{t}}^{2} $$

$$ - 3{K_{11}}^{-2}{K_{33}}^{-1}h{K_{33}},_{t} -
{K_{11}}^{-2}{K_{33}}^{-1}{K_{33}},_{t t} + {1 \over
2}{K_{11}}^{-2}{K_{33}}^{2}hf,_{t}g,_{t} $$

$$ - {1 \over 4}{K_{11}}^{-2}{K_{33}}^{2}h^{2}{f,_{t}}^{2} - {1 \over
4}{K_{11}}^{-2}{K_{33}}^{2}{g,_{t}}^{2} - 3{K_{11}}^{-2}h^{2} - {5 \over
2}{K_{11}}^{-2}h,_{t} $$

$$ + {K_{11}}^{-1}K^{-1}{K_{11}},_{t}K,_{t} +
{K_{11}}^{-1}{K_{33}}^{-1}{K_{11}},_{t}{K_{33}},_{t} - {1 \over
4}K^{-2}{K_{33}}^{2}{f,_{t}}^{2} $$

$$ + K^{-1}{K_{33}}^{-1}K,_{t}{K_{33}},_{t} - 3K^{-2} \eqno{(A.1)} $$

\bigskip

$$ G_{01} = - 2{K_{11}}^{-2}h{K_{11}},_{t} -
{K_{11}}^{-2}K^{-1}{K_{11}},_{t}K,_{t} -
{K_{11}}^{-2}{K_{33}}^{-1}{K_{11}},_{t}{K_{33}},_{t} $$

$$ + {1 \over 2}{K_{11}}^{-1}K^{-2} + {1 \over
2}{K_{11}}^{-1}K^{-2}{K_{33}}^{2}{f,_{t}}^{2} + {K_{11}}^{-1}K^{-1}hK,_{t} $$

$$ + {K_{11}}^{-1}K^{-1}K,_{t t} + {K_{11}}^{-1}{K_{33}}^{-1}h{K_{33}},_{t} +
{K_{11}}^{-1}{K_{33}}^{-1}{K_{33}},_{t t} + {3 \over 2}{K_{11}}^{-1}h,_{t}
\eqno{(A.2)} $$

\bigskip

$$ G_{02} = {1 \over 2}{K_{11}}^{-3}K{K_{11}},_{t}h,_{t} - {1 \over
2}{K_{11}}^{-3}K^{-1}{K_{11}},_{t} - {3 \over 2}{K_{11}}^{-2}Khh,_{t} $$

$$ - {1 \over 2}{K_{11}}^{-2}K{K_{33}}^{-1}{K_{33}},_{t}h,_{t} - {1 \over
2}{K_{11}}^{-2}Kh,_{t t} - {1 \over 2}{K_{11}}^{-2}K^{-2}K,_{t} $$

$$ + {1 \over 2}{K_{11}}^{-2}K^{-1}h + {1 \over
2}{K_{11}}^{-2}K^{-1}{K_{33}}^{-1}{K_{33}},_{t} + {1 \over
2}{K_{11}}^{-2}K^{-1}{K_{33}}^{2}h{f,_{t}}^{2} $$

$$ - {1 \over 2}{K_{11}}^{-2}K^{-1}{K_{33}}^{2}f,_{t}g,_{t} - {3 \over
2}{K_{11}}^{-2}K,_{t}h,_{t} - {K_{11}}^{-1}K^{-1}{K_{11}},_{t} $$

$$ + 2K^{-2}K,_{t} - K^{-1}{K_{33}}^{-1}{K_{33}},_{t} \eqno{(A.3)} $$

\bigskip

$$ G_{03} = - {1 \over 2}{K_{11}}^{-3}{K_{33}}h{K_{11}},_{t}f,_{t} + {1 \over
2}{K_{11}}^{-3}{K_{33}}{K_{11}},_{t}g,_{t} + {1 \over
2}{K_{11}}^{-2}{K_{33}}hf,_{t t} $$

$$ - {3 \over 2}{K_{11}}^{-2}{K_{33}}hg,_{t} + {3 \over
2}{K_{11}}^{-2}{K_{33}}h^{2}f,_{t} - {1 \over 2}{K_{11}}^{-2}{K_{33}}g,_{t t} +
{1 \over 2}{K_{11}}^{-2}{K_{33}}f,_{t}h,_{t} $$

$$ + {3 \over 2}{K_{11}}^{-2}h{K_{33}},_{t}f,_{t} - {1 \over
2}{K_{11}}^{-2}K^{-2}{K_{33}}f,_{t} + {1 \over
2}{K_{11}}^{-2}K^{-1}{K_{33}}hK,_{t}f,_{t} $$

$$ - {1 \over 2}{K_{11}}^{-2}K^{-1}{K_{33}}K,_{t}g,_{t} - {3 \over
2}{K_{11}}^{-2}{K_{33}},_{t}g,_{t} + {3 \over 2}K^{-2}{K_{33}}f,_{t}
\eqno{(A.4)} $$

\bigskip

$$ G_{11} = {1 \over 4}{K_{11}}^{-2}K^{-2} - {1 \over
4}{K_{11}}^{-2}K^{-2}{K_{33}}^{2}{f,_{t}}^{2} + {K_{11}}^{-2}K^{-1}hK,_{t} $$

$$ + {K_{11}}^{-2}K^{-1}{K_{33}}^{-1}K,_{t}{K_{33}},_{t} + {1 \over
4}{K_{11}}^{-2}K^{2}{h,_{t}}^{2} + {K_{11}}^{-2}{K_{33}}^{-1}h{K_{33}},_{t} $$

$$ - {1 \over 2}{K_{11}}^{-2}{K_{33}}^{2}hf,_{t}g,_{t} + {1 \over
4}{K_{11}}^{-2}{K_{33}}^{2}h^{2}{f,_{t}}^{2} + {1 \over
4}{K_{11}}^{-2}{K_{33}}^{2}{g,_{t}}^{2} $$

$$ + {K_{11}}^{-2}h^{2} - {1 \over 2}{K_{11}}^{-2}h,_{t} - {1 \over
4}K^{-2}{K_{33}}^{2}{f,_{t}}^{2} - K^{-1}{K_{33}}^{-1}K,_{t}{K_{33}},_{t} $$

$$ - K^{-1}K,_{t t} - {K_{33}}^{-1}{K_{33}},_{t t} + K^{-2} \eqno{(A.5)} $$

\bigskip

$$ G_{12} = - {1 \over 2}{K_{11}}^{-2}K{K_{11}},_{t}h,_{t} + {1 \over
2}{K_{11}}^{-2}K^{-1}{K_{11}},_{t} + {1 \over
2}{K_{11}}^{-1}K{K_{33}}^{-1}{K_{33}},_{t}h,_{t} $$

$$ + {1 \over 2}{K_{11}}^{-1}Kh,_{t t} - {3 \over 2}{K_{11}}^{-1}K^{-2}K,_{t} +
{1 \over 2}{K_{11}}^{-1}K^{-1}{K_{33}}^{-1}{K_{33}},_{t} $$

$$ - {1 \over 2}{K_{11}}^{-1}K^{-1}{K_{33}}^{2}h{f,_{t}}^{2} + {1 \over
2}{K_{11}}^{-1}K^{-1}{K_{33}}^{2}f,_{t}g,_{t} + {3 \over
2}{K_{11}}^{-1}K,_{t}h,_{t} \eqno{(A.6)} $$

\bigskip

$$ G_{13} = {1 \over 2}{K_{11}}^{-2}{K_{33}}h{K_{11}},_{t}f,_{t} - {1 \over
2}{K_{11}}^{-2}{K_{33}}{K_{11}},_{t}g,_{t} - {1 \over
2}{K_{11}}^{-1}{K_{33}}hf,_{t t} $$

$$ + {1 \over 2}{K_{11}}^{-1}{K_{33}}g,_{t t} - {1 \over 2}{K_{11}}
^{-1}{K_{33}}f,_{t}h,_{t} - {3 \over 2}{K_{11}}^{-1}h{K_{33}},_{t}f,_{t} -
{K_{11}}^{-1}K^{-2}{K_{33}}f,_{t} $$

$$ - {1 \over 2}{K_{11}}^{-1}K^{-1}{K_{33}}hK,_{t}f,_{t} + {1 \over
2}{K_{11}}^{-1}K^{-1}{K_{33}}K,_{t}g,_{t} + {3 \over
2}{K_{11}}^{-1}{K_{33}},_{t}g,_{t} \eqno{(A.7)} $$

\bigskip

$$ G_{22} = - {K_{11}}^{-3}h{K_{11}},_{t} - {K_{11}}^{-3}{K_{33}}^{-1}
{K_{11}},_{t}{K_{33}},_{t} + {1 \over 4}{K_{11}}^{-2}K^{-2} - {1 \over
4}{K_{11}}^{-2}K^{-2}{K_{33}}^{2}{f,_{t}}^{2} $$

$$ - {3 \over 4}{K_{11}}^{-2}K^{2}{h,_{t}}^{2} + 2{K_{11}}^{-2}{K_{33}}^{
-1}h{K_{33}},_{t} + {K_{11}}^{-2}{K_{33}}^{-1}{K_{33}},_{t t} + {1 \over
2}{K_{11}}^{-2}{K_{33}}^{2}hf,_{t}g,_{t} $$

$$ - {1 \over 4}{K_{11}}^{-2}{K_{33}}^{2}h^{2}{f,_{t}}^{2} - {1 \over 4}{K_{
11}}^{-2}{K_{33}}^{2}{g,_{t}}^{2} + {K_{11}}^{-2}h^{2} + {3 \over
2}{K_{11}}^{-2}h,_{t} $$

$$ - {K_{11}}^{-1}{K_{33}}^{-1}{K_{11}},_{t}{K_{33}},_{t} -
{K_{11}}^{-1}{K_{11}},_{t t} + {1 \over 4}K^{-2}{K_{33}}^{2}{f,_{t}}^{2} -
{K_{33}}^{-1}{K_{33}},_{t t} + K^{-2} \eqno{(A.8)} $$

\bigskip

$$ G_{23} = {1 \over 2}{K_{11}}^{-3}K^{-1}{K_{33}}{K_{11}},_{t}f,_{t} + {1
\over 2}{K_{11}}^{-2}K{K_{33}}hf,_{t}h,_{t} - {1 \over
2}{K_{11}}^{-2}K{K_{33}}g,_{t}h,_{t} $$

$$ + {1 \over 2}{K_{11}}^{-2}K^{-2}{K_{33}}K,_{t}f,_{t} -
{K_{11}}^{-2}K^{-1}{K_{33}}hf,_{t} - {1 \over
2}{K_{11}}^{-2}K^{-1}{K_{33}}f,_{t t} $$

$$ - {3 \over 2}{K_{11}}^{-2}K^{-1}{K_{33}},_{t}f,_{t} + {1 \over
2}{K_{11}}^{-1}K^{-1}{K_{33}}{K_{11}},_{t}f,_{t} - {1 \over
2}K^{-2}{K_{33}}K,_{t}f,_{t} $$

$$ + {1 \over 2}K^{-1}{K_{33}}f,_{t t} + {3 \over 2}K^{-1}{K_{33}},_{t}f,_{t}
\eqno{(A.9)} $$

\bigskip

$$ G_{33} = - {K_{11}}^{-3}h{K_{11}},_{t} -
{K_{11}}^{-3}K^{-1}{K_{11}},_{t}K,_{t} - {1 \over 4}{K_{11}}^{-2}K^{-2} + {3
\over 4}{K_{11}}^{-2}K^{-2}{K_{33}}^{2}{f,_{t}}^{2} $$

$$ + 2{K_{11}}^{-2}K^{-1}hK,_{t} + {K_{11}}^{-2}K^{-1}K,_{t t} - {1 \over
4}{K_{11}}^{-2}K^{2}{h,_{t}}^{2} + {3 \over
2}{K_{11}}^{-2}{K_{33}}^{2}hf,_{t}g,_{t} $$

$$ - {3 \over 4}{K_{11}}^{-2}{K_{33}}^{2}h^{2}{f,_{t}}^{2} - {3 \over
4}{K_{11}}^{-2}{K_{33}}^{2}{g,_{t}}^{2} + {K_{11}}^{-2}h^{2} + {3 \over
2}{K_{11}}^{-2}h,_{t} - {K_{11}}^{-1}K^{-1}{K_{11}},_{t}K,_{t} $$

$$ - {K_{11}}^{-1}{K_{11}},_{t t} - {3 \over 4}K^{-2}{K_{33}}^{2}{f,_{t}}^{2} -
K^{-1}K,_{t t} + K^{-2} \eqno{(A.10)} $$

\noindent Since the source in the Einstein equations is dust with a
cosmological constant, and since the zero-th tetrad vector is just the velocity
vector, the above components should obey the following equations:

$$
G_{00} = (8\pi G/c^4) \rho - \Lambda,
$$

$$
G_{11} = G_{22} = G_{33} = \Lambda,
$$

$$
{\rm nondiagonal} \qquad G_{IJ} = 0, \eqno{(A.11)} $$

\noindent where $\rho$ is the dust energy-density and $\Lambda$ is the
cosmological constant.

\bigskip

\begin{center}

{\bf Appendix B}

{\bf Consequences of $h = 0$ in the Einstein equations.}

\end {center}

With $h = 0$, eq. (2.12) becomes:

$$
K_{33}f,_t({K_{11}}^2 - 1) / (K_{11}K^2) = 0 \eqno {(B.1)}
$$

\noindent We can immediately discard the solution $K_{33} = 0$ because then
$\det(g_{\alpha \beta}) = 0$. When ${K_{11}}^2 = 1$, the limit $\omega \to 0$
of the resulting metric will necessarily have either nonzero shear or zero
expansion (see sec. 3 for a calculation of this limit), and so no
generalization of the Friedmann models can be expected here. Hence, the only
consequence of (B.1) that is worth pursuing is:

$$
f,_t = 0. \eqno{(B.2)}
$$

\noindent Then, with $h = 0$, eq. (A.6) implies:

$$
K_{11}K_{33}/K^3 = {\rm const}. \eqno {(B.3)}
$$

\noindent However, in the limit $\omega \to 0$ this again implies either
nonzero shear or zero expansion, i.e. no Friedmann limit.

\bigskip

\begin{center}

{\bf Appendix C}

{\bf The result for $F,_{tt}$ from $G_{01} = 0$.}

\end {center}

When (4.2), (4.2) and (4.3) are substituted in (A.2), the following formula
results for $F,_{tt}$:

$$ F,_{t t} = - {3 \over 2}Kh{\cosh}^{-1}(F){\sinh}^{-1}(F)K,_{t}h,_{t} +
3h{\cosh}^{-2}(F)F,_{t} - {7 \over 2}h{\sinh}^{-2}(F)F,_{t} - 3hF,_{t} $$

$$ + {1 \over 2}K^{-2}{\cosh}^{-1}(F)\sinh (F) + 2K^{-2}{\cosh}^{-1}(F)\sinh
(F){K,_{t}}^{2} $$

$$ + {1 \over 2}K^{-1}h{\cosh}^{-1}(F){\sinh}^{-1}(F)K,_{t}+
3K^{-1}h{\cosh}^{-1}(F)\sinh (F)K,_{t} $$

$$ + 3K^{-1}{\cosh}^{-2}(F)K,_{t}F,_{t} + 3K^{-1}{\cosh}^{-1}(F)\sinh (F)K,_{t
t} - 7K^{-1}K,_{t}F,_{t} $$

$$ + {9 \over 2}K^{2}h{\sinh}^{-2}(F)F,_{t}h,_{t}-
3K^{2}h^{2}{\cosh}^{-1}(F){\sinh}^{-3}(F)h,_{t} $$

$$ - {3 \over 2}K^{2}h^{2}{\cosh}^{-1}(F){\sinh}^{-1}(F)h,_{t}- {3 \over
2}K^{2}{\cosh}^{-1}(F){\sinh}^{-1}(F){h,_{t}}^{2} $$

$$ + h^{2}{\cosh}^{-1}(F){\sinh}^{-3}(F) + {1 \over
2}h^{2}{\cosh}^{-1}(F){\sinh}^{-1}(F) +
2{\cosh}^{-1}(F){\sinh}^{-1}(F){F,_{t}}^{2} $$

$$ + {1 \over 2}{\cosh}^{-1}(F){\sinh}^{-1}(F)h,_{t} + 2{\cosh}^{-1}(F)\sinh
(F){F,_{t}}^{2} + {3 \over 2}{\cosh}^{-1}(F)\sinh (F)h,_{t}. $$

\noindent It will be modified later because it contains $K,_{tt}$ on the
right-hand side, while the final equations that will be dealt with should have
no second derivatives on the right-hand sides.

\begin{center}

{\bf Appendix D}

{\bf Nonexistence of a Hilbert-type Lagrangian for the set \{(4.3), (4.5),
(4.6)\}.}

\end {center}

Eqs. (4.3), (4.5) and (4.6) can be written in the form:

$$ {{{\rm d}^2z^i} \over {{\rm d}t^2}} = {W^i}_{jk}{{{\rm d}z^j} \over {{\rm
d}t}} {{{\rm d}z^k} \over {{\rm d}t}} + {V^i}_j{{{\rm d}z^j} \over {{\rm d}t}}
+ U^i, \eqno{(D.1)} $$

\noindent where $i = 0, 1, 2$; $z^0 = h$, $z^1 = K$, $z^2 = F$ and
${W^i}_{jk}$, ${V^i}_j$ and $U^i$ are functions of $(h, K, F)$ (but not of
$t$). Note that the set (D.1) is covariant with respect to arbitrary
transformations $z^i \to z'^i = f^i(\{z^j\})$: the first derivatives ${{{\rm
d}z^j} \over {{\rm d}t}}$ transform then like a contravariant vector, and so do
the terms $U^i$, the coefficients ${V^i}_j$ transform like a mixed tensor, and
the coefficients $(- {W^i}_{jk})$ transform like components of an affine
connection. (The nontensorial terms in the transformed $(- {W^i}_{jk})$ arise
from ${{{\rm d}^2z^i} \over {{\rm d}t^2}}$). The most natural ansatz for a
lagrangian for (D.1) is:

$$ L = Q_{ij} {{{\rm d}z^i} \over {{\rm d}t}} {{{\rm d}z^j} \over {{\rm d}t}} +
L_i {{{\rm d}z^i} \over {{\rm d}t}} + \Phi, \eqno{(D.2)} $$

\noindent where $Q_{ij}$, $L_i$ and $\Phi$ are functions of $(h, K, F)$. Such a
lagrangian would result from the Hilbert lagrangian by taking out a complete
divergence and integrating the result with respect to the spatial variables.
The Euler-Lagrange equations implied by (D.2) are:

$$ Q_{is} {{{\rm d}^2z^s} \over {{\rm d}t^2}} = - (Q_{ki,l} - {1 \over 2}
Q_{kl,i}) {{{\rm d}z^k} \over {{\rm d}t}} {{{\rm d}z^l} \over {{\rm d}t}} + {1
\over 2}(L_{k,i} - L_{i,k}){{{\rm d}z^k} \over {{\rm d}t}} + {1 \over 2}\Phi,_i
\eqno{(D.3)} $$

\noindent If these are to be equivalent to (D.1), then the following must hold:

$$
Q_{is}{W^s}_{kl} = - {1 \over 2}(Q_{ki,l} + Q_{li,k} - Q_{kl,i}), \eqno{(D.4)}
$$

$$
Q_{is}{V^s}_{k} = {1 \over 2}(L_{k,i} - L_{i,k}), \eqno{(D.5)}
$$

$$
Q_{is}U^s = {1 \over 2}\Phi,_i. \eqno{(D.6)}
$$

\noindent Eqs. (D.4) imply that $(- {W^i}_{jk})$ must be Christoffel symbols
constructed from the metric $Q_{ij}$, eqs. (D.5) imply that ${1 \over 2}L_i$
must be a vector potential for the tensor field $Q_{is}{V^s}_k$, and eqs. (D.6)
imply that $\Phi /2$ must be a scalar potential for the vector field
$Q_{is}U^s$. All of these are strong conditions and they may be impossible to
fulfil in many cases.

Indeed, for our eqs. \{(4.3), (4.5), (4.6)\}, the solution of (D.4) turns out
to be $Q_{ij} \equiv 0$, i.e. the Lagrangian (D.2) does not exist. This is an
outline of the proof.

After eqs. (D.4) are written out in the form

$$
Q_{ij,k} = - {W^s}_{ik}Q_{sj} - {W^s}_{jk}Q_{is}, \eqno{(D.7)}
$$

\noindent with ${W^s}_{kl}$ read off from \{(4.3), (4.5), (4.6)\}, the
following two equations follow, among other results:

$$ Q_{11,F} + {1 \over 4}(K \cosh(F)/ \sinh(F)) Q_{11,K} = - (2 {\cosh}^2(F) -
1)Q_{11}/(\cosh(F) \sinh(F)), \eqno{(D.8)} $$

$$ Q_{22,K} + (2 {\cosh}^2(F) - 1) \sinh(F)Q_{22,F} /(2K {\cosh}^3(F)) = (3
{\cosh}^2(F) - 1)Q_{22}/(K{\cosh}^2(F)). \eqno{(D.9)} $$

\noindent The solutions of these are:

$$ Q_{11} = q_{11}\left( h, {{K^4} \over {\sinh(F)}}\right) {1 \over {\cosh(F)
\sinh(F)}}, \eqno{(D.10)} $$

$$ Q_{22} = K{\sinh}^2(F) q_{22}\left( h, {{K{\sqrt {2 {\cosh}^2(F) - 1}}}
\over {{\sinh}^2(F)}}\right) , \eqno{(D.11)} $$

\noindent where $q_{ij}$ are arbitrary functions of their two arguments.
The equation $Q_{11,K} = \dots$ is then solved with the result:

$$
Q_{12} = - K^5q_{11,w}/{\sinh}^3(F), \eqno{(D.12)}
$$

\noindent where $w = K^4/ \sinh(F)$ is the second argument of $q_{11}$, and the
equation $Q_{22,K} = \dots$ implies:

$$
{1 \over v}q_{22,v} = q_{11,w}K^3/(\cosh(F){\sinh}^2(F)), \eqno{(D.13)}
$$

\noindent where $v$ is the second argument of $q_{22}$. The left-hand side of
(D.13) is an invariant of the operator $(2K{\cosh}^3(F)/ \sinh(F)){{\partial}
\over {\partial K} } + (2 {\cosh}^2(F) - 1){{\partial} \over {\partial F} }$,
while $q_{11,w}$ is an invariant of the operator ${1 \over 4}K{{\partial} \over
{\partial K} } + (\sinh(F)/ \cosh(F)){{\partial} \over {\partial F} }$.
Application of these two operators to (D.13) leads to $q_{22,v} = q_{11,w} =
0$, which implies $Q_{12} = 0$. With this, the remaining equations (D.7)
quickly lead to $Q_{ij} \equiv 0$, which means that the Lagrangian (D.2) does
not exist in this case.

Since the Euler-Lagrange equations (D.4) are covariant with respect to
arbitrary transformations of the Lagrangian variables (in our case $h \to h'(h,
K, F)$, etc.), and equations of the form (D.1) are covariant, too, the
conclusion that a Lagrangian of the form (D.2) exists (or does not exist) is
coordinate-independent, i.e. having shown that eqs. $\{(4.3), (4.5), (4.6)\}$
do not follow from a Lagrangian (D.2) in our variables $\{h, K, F\}$, we know
that no such Lagrangian will exist in any other variables.

\begin{center}

{\bf Appendix E}

{\bf The general solution of eqs. (5.8) -- (5.11).}

\end {center}

Eqs. (5.8) have the form:

$$
\xi_{;kl} = 0, \eqno{(E.1)}
$$

\noindent where ; is the covariant derivative in which $(- {W^i}_{kl})$ play
the role of the connection coefficients. (They appear in this role for a second
time already, see Appendix D.) The integrability conditions of (E.1) are:

$$
{R^s}_{ijk}\xi,_s = 0, \eqno{(E.2)}
$$

\noindent where ${R^s}_{ijk} = - {R^s}_{ikj}$ is the curvature tensor
corresponding to the connection$(- {W^i}_{kl})$. Eqs. (E.2) are 9 equations
(labelled by the sets of indices $(i, j, k) = (0, 0, 1); (0, 0, 2); (0, 1, 2)$;
etc) and they could have nontrivial solutions only if every subset of 3
equations chosen from among them had a zero determinant. Actually, of the 84
determinants only two vanish, and some of them will not vanish even if the
functions $h(t)$, $K(t)$ and $F(t)$ are functionally dependent. Here is one
determinant that will never vanish, it corresponds to $\{(i, j, k)\} = \{(1, 0,
1); (1, 1, 2); (2, 1, 2)\}$:

$$ {\rm det}(E.2) = K^{-4}[{(- {189 \over 32} + {3 \over 4}{\cosh}^{-6}(F) +
{53 \over 16}{\cosh}^{-4}(F) + {5 \over 8}{\cosh}^{-2}(F) - {11 \over
8}{\sinh}^{-2}(F))}]. $$

\noindent Hence, the unique solution of (5.8) is:

$$
\xi = \xi(t). \eqno{(E.3)}
$$

With $\xi,_i = 0$, eqs. (5.9) simplify somewhat, and the equation corresponding
to $(i, k, l) = (0, 1, 2)$ becomes ${\eta^0},_{K F} = - 2 {\eta^0},_F / K$,
which has the solution:

$$
\eta^0 = F^0(t, h, F)/K^2 + G^0(t, h, K), \eqno{(E.4)}
$$

\noindent $F^0$ and $G^0$ being unknown functions. Then, eq. (5.9) with $(i, k,
l) = (0, 1, 1)$ allows us to separate the variables $K$ and $F$, and its
solution, substituted into (E.4), gives the result:

$$
\eta^0 = M^0(t, h){\sinh}^2(F)/K^2 + J^0(t, h)/K^4 + L^0(t, h), \eqno{(E.5)}
$$

\noindent where $M^0$, $J^0$ and $L^0$ are new unknown functions. With this,
eq. (5.9) corresponding to $(i, k, l) = (0, 2, 2)$ implies $J^0 = 0$, and the
one with $(i, k, l) = (0, 0, 2)$ solves as follows:

$$ \eta^1 = {3 \over 5}M^0 Kh \log ({\sinh}(F)) - {2 \over {5K}}
{M^0},_h{\sinh}^2(F) + {{K(2{\cosh}^2(F) - 1)} \over {5 \cosh(F)
\sinh(F)}}\eta^2 + F^1(t, h, K), \eqno{(E.6)} $$

\noindent where $F^1(t, h, K)$ is a new unknown function, and $\eta^2$ is still
completely unknown. Then, for $(i, k, l) = (0, 0, 1)$, the equation (5.9) has
the solution:

$$
F^1 = -{3 \over 5}M^0 Kh \log K + G^1(t, h) K, \eqno{(E.7)}
$$

\noindent where $G^1(t, h)$ is a new unknown function.

When (E.6) and (E.7) are substituted into the (0, 0, 0) component of (5.9), an
algebraic equation for $\eta^2$ results, whose solution is:

$$ \eta^2 = {1 \over {3\cosh^2(F) + 1}} \left\{ {5 \over 3} {{{M^0},_{hh}}
\over {K^4h}} \cosh(F) \sinh^5(F) - {5 \over 3} {{{L^0},_{hh}} \over {K^2h}}
\cosh(F) \sinh^3(F) \right. $$

$$ + {35 \over 6} {{M^0} \over {K^2h}} \cosh(F) \sinh^3(F) + {5 \over 2} {{L^0}
\over h} \cosh(F) \sinh(F) $$

$$
+ 3M^0h\cosh(F) \sinh(F) [\log (\sinh(F)) - \log K] + 5\cosh(F) \sinh(F)G^1
$$

$$ + {1 \over {2K^2}} {M^0},_h \cosh(F) \sinh^3(F) + {5 \over 2} {L^0},_h
\cosh(F) \sinh(F) $$

$$ \left. - 5 {{M^0} \over {K^2h}} \cosh(F) \sinh^3(F) [\log (\sinh(F)) - \log
K] - {25 \over 3} {{{G^1},_h} \over {K^2h}} \cosh(F) \sinh^3(F) \right\} .
\eqno{(E.8)} $$

\noindent Both sides of the (1, 1, 1) component of (5.9) become then
polynomials in $(\log K)$ and $1/K$, whose corresponding coefficients have to
be respectively equal. The coefficients of $K^{-1} \log K$ imply then $M^0 =
0$, and with this, only two other terms remain whose solutions are:

$$
L^0 = 3C(t)h^{-5/3} - B(t) h, \qquad G^1 = C(t) h^{-8/3} + B(t). \eqno{(E.9)}
$$

\noindent where $C(t)$ and $B(t)$ are unknown functions. But this results in
$\eta^2 = 0$, $\eta^0 = L^0$, $\eta^1 = G^1 K$. Then, the (1, 0, 0) component
of (5.9) implies $C(t) = 0$, and with this all the remaining equations in (5.9)
are fulfilled. Thus the final solution of (5.9) is:

$$
\eta^0 = - B(t)h, \qquad \eta^1 = B(t)K, \qquad \eta^2 = 0. \eqno{(E.10)}
$$

\noindent With $\xi = \xi(t)$ from (E.3) and $\eta^i$ as above, any equation of
the set (5.10) implies:

$$
B = {\rm const}, \qquad \xi = Bt +A, \qquad A = {\rm const}, \eqno{(E.11)}
$$

\noindent and this satisfies all the remaining equations (5.10) and (5.11).
Hence, the general solution of (5.8) -- (5.11) is (5.12).

This result was derived under the tacit assumption that the functions $h$, $K$
and $F$ are functionally independent. In the course of solving the equations
(4.3), (4.5) and (4.6), relations between these functions may appear. It
happens sometimes that such relations are revealed by the symmetry equations as
cases in which the symmetry group is larger than in the generic case (see e.g.
Ref. 16 where special cases of larger symmetry of a single equation were
revealed by the symmetry equations). This possibility has not been investigated
for the equations (5.9) -- (5.11). However, for the equation (5.8) the solution
is always (E.3), even if the functions $h$, $K$ and $F$ are not independent, as
shown in the paragraph containing (E.3).

{\Large {\bf References.}}

$^1$ A. Krasi\'nski, {\it J. Math. Phys.} {\bf 39}, 380 (1998).

$^2$ A. Krasi\'nski, {\it J. Math. Phys.} {\bf 39}, 401 (1998).

$^3$ A. Krasi\'nski, {\it J. Math. Phys.} {\bf 39}, 2148 (1998).

$^4$ A. Krasi\'nski, {\it Solutions of the Einstein field equations for a
rotating perfect fluid. Part 3: A survey of models of rotating perfect fluid or
dust.} Preprint, Warsaw 1975.

$^5$ A. Krasi\'nski, {\it Inhomogeneous cosmological models.} Cambridge
Unviversity Press 1997.

$^6$ P. Szekeres, {\it Commun. Math. Phys. } {\bf 41}, 55 (1975).

$^7$ D. A. Szafron, {\it J. Math. Phys.} {\bf 18}, 1673 (1977).

$^8$ C. Hellaby, {\it Class. Quant. Grav.} {\bf 13}, 2537 (1996).

$^9$ H. Stephani, {\it Differential equations; their solution using
symmetries}. Cambridge University Press 1989.

$^{10}$ P. J. Olver, {\it Applications of Lie groups to differential
equations}, 2nd edition. Springer 1993, pp. 350 -- 379.

$^{11}$ M. A. H. Mac Callum, in {\it General relativity, an Einstein centenary
survey}. Edited by S. W. Hawking and W. Israel. Cambridge University Press
1979, p. 552 -- 553.

$^{12}$ G. F. R. Ellis, {\it J. Math. Phys.} {\bf 8}, 1171 (1967).

$^{13}$ A. Krasi\'nski, {\it Gen. Rel. Grav.} {\bf 25}, 165 (1993).

$^{14}$ A. Krasi\'nski, {\it The newest release of the Ortocartan set of
programs for algebraic calculations in relativity}. Submitted for publication.

$^{15}$ A. Krasi\'nski, M. Perkowski, {\it The system Ortocartan -- user's
manual}. Fifth edition, 2000. Available by email or on a diskette.

$^{16}$ H. Stephani, {\it J. Phys.} {\bf A16}, 3529 (1983).

\end{document}